\documentclass{IOS-Book-Article}
\usepackage{mathptmx}
\usepackage{xspace}
\usepackage{url}
\usepackage{color}
\usepackage{multirow}
\usepackage{graphicx}

\newcommand{\CPTWOK}{{\tt CP2K}\xspace}
\newcommand{\DBCSR}{{\tt DBCSR}\xspace}
\newcommand{\LIBXSMM}{{\tt LIBXSMM}\xspace}
\newcommand{\LIBCUSMM}{{\tt LIBCUSMM}\xspace}
\newcommand{\MKL}{{\tt MKL}\xspace}

\newcommand{\IE}{{i.\,e.}\xspace}

\def\hb{\hbox to 10.7 cm{}}

\begin{document}

\pagestyle{headings}
\def\thepage{}

\begin{frontmatter}              

\title{Porting of the \DBCSR library for Sparse Matrix-Matrix Multiplications to Intel Xeon Phi systems}

\markboth{}{August 2017\hb}


\author[A]{\fnms{Iain} \snm{Bethune}\thanks{E-mail: iain.bethune@stfc.ac.uk}},
\author[B]{\fnms{Andreas} \snm{Gl{\"o}ss}\thanks{E-mail: andreas.gloess@chem.uzh.ch}},
\author[B]{\fnms{J{\"u}rg} \snm{Hutter}\thanks{E-mail: hutter@chem.uzh.ch}},
\author[B]{\fnms{Alfio} \snm{Lazzaro}\thanks{E-mail: alfio.lazzaro@chem.uzh.ch}},
\author[C]{\fnms{Hans} \snm{Pabst}\thanks{E-mail: hans.pabst@intel.com}}
and
\author[D]{\fnms{Fiona} \snm{Reid}\thanks{E-mail: f.reid@epcc.ed.ac.uk}}

\address[A]{Hartree Centre, Science and Technology Facilities Council, United Kingdom}
\address[B]{University of Zurich, Department of Chemistry, Switzerland}
\address[C]{Intel Semiconductor AG, Switzerland}
\address[D]{EPCC, The University of Edinburgh, United Kingdom}

\begin{abstract}
Multiplication of two sparse matrices is a key operation in the simulation of the electronic structure of systems containing thousands of atoms and electrons. The highly optimized sparse linear algebra library \DBCSR (Distributed Block Compressed Sparse Row) has been specifically designed to efficiently perform such sparse matrix-matrix multiplications. This library is the
basic building block for linear scaling electronic structure theory and low scaling correlated methods in \CPTWOK. It is parallelized using MPI and OpenMP, and can exploit GPU accelerators by means of CUDA.
We describe a performance comparison
of \DBCSR on systems with Intel Xeon Phi Knights Landing (KNL) processors,
with respect to systems with Intel Xeon CPUs (including systems with GPUs). 

We find that the \DBCSR on Cray XC40 KNL-based systems is 11\%-14\% slower than on a hybrid Cray XC50 with Nvidia P100 cards, at the same
number of nodes.  When compared to a Cray XC40 system equipped with dual-socket Intel Xeon CPUs, the KNL is up to 24\% faster.
\end{abstract}

\end{frontmatter}
\markboth{August 2017\hb}{August 2017\hb}

\section{Introduction}

Multiplication of two sparse matrices (SpGEMM) is a key operation in the simulation of the electronic structure of systems containing thousands of atoms and electrons~\cite{joost1M}. Examples of such systems include electronic devices, complex interfaces, macromolecules and large disordered systems, with applications in the fields of renewable energy and electronics. The theory that enables such studies is linear scaling Density Functional Theory (DFT)~\cite{LS}. 
In the atomistic simulation package \CPTWOK~\cite{cp2k}, the linear scaling DFT implementation exploits the fact that for large enough systems, operators in a localized atomic basis become sparse~\cite{joost1M}. The matrices have several thousands of non-zero elements per row and \emph{a priori} unknown sparsity patterns.
In these simulations, SpGEMM typically accounts for more than 80\% of the total runtime.
The computational cost depends strongly on the evolution of the sparsity during the iterations,
which in turn depends on the chemical properties of the system studied, the precise algorithm employed, the system size, and the required accuracy~\cite{joost1M}.
The highly optimized sparse linear algebra library \DBCSR (Distributed Block Compressed Sparse Row) has been specifically designed to efficiently perform such block-sparse matrix-matrix multiplications~\cite{dbcsr, ole, Lazzaro:2017:IES:3093172.3093228}. It is parallelized using MPI and OpenMP, and can exploit GPU accelerators by means of CUDA. 

Here we describe our evaluation of \DBCSR on systems equipped with Intel Xeon Phi `Knights Landing' (KNL) processors. These systems are emerging as a viable but segment-specific alternative to traditional x86-64 CPU systems and systems with GPUs to reach higher computational density~\cite{top500}. Although running on the Intel Xeon Phi is straightforward, it poses several challenges for the application in order to obtain good performance, such as vectorization, memory management and multi-threading~\cite{portingKNC,optimisingKNC}. 
We compare performance
between runs on KNL systems with respect to systems with Intel Xeon CPUs (including systems with GPUs), on up to 144 nodes.  
In the interest of portability, the same \DBCSR code was used for CPU and KNL executions, \IE we do not use any particular code optimization specific for KNL systems besides the optimization provided by the compiler.

\subsection{Related Work}

The classical serial SpGEMM algorithm was first described by Gustavson~\cite{Gustavson:1978:TFA:355791.355796}.
The parallel implementation in a distributed memory system presents several challenges, such as load-balance and 
communication costs relative to arithmetic operations, and several algorithms have been proposed~\cite{10.1109/ICPP.2008.45,ibm,
Ballard:2016:HPS:3012407.3015144}. 
\DBCSR considers the general case
where \emph{a priori} knowledge of the input and output matrix sparsity is not employed, and is aimed at delivering good performance in the `nearly dense' regime \IE many non-zeros per row. It uses a random permutation of the rows and columns of the matrix to achieve a good average load-balance. Consequently, the data and the corresponding operations are statically distributed across processes in the same way as for dense matrices, and existing algorithms for dense matrix-matrix multiplications (e.g. ~\cite{Ballard:2013:COP:2486159.2486196}) can be adopted and refined for the sparse case.
Recently, we have implemented a 2.5D algorithm that is able to improve the performance for large number of processors~\cite{Lazzaro:2017:IES:3093172.3093228}.

Several papers report on SpGEMM implementations for single-node GPU-enabled systems~\cite{Dalton:2015:OSM:2835205.2699470, journals/siamsc/GremseHSKN15, Polok:2015:FSM:2872599.2872604}. The work of Liu and Vinter~\cite{Liu201547} addresses heterogeneous CPU-GPU processors. A recent paper by Deveci {\it et al.}~\cite{deveci2017performance} introduces an implementation that can target both GPU and KNL. None of these implementations target block-sparse matrices.
Concerning parallel implementation in a hybrid CPU-GPU multinode system, Rubensson and Rudberg~\cite{rubensson_rudberg} reported a parallel implementation where the
mapping of data and work to physical resources is performed dynamically during the calculation.
Like \DBCSR, this implementation is able to work effectively with block-sparse matrices and runs on hybrid multi-cores CPU and GPU systems, 
however it does not employ optimized libraries for the small block multiplications.

\section{{\large \textbf{\DBCSR}} Library}
\label{sec:DBCSR}

\DBCSR is written in Fortran and is freely available under the GPL license from \url{https://dbcsr.cp2k.org}.
\DBCSR matrices are stored in a blocked compressed sparse row (CSR) format distributed over a two-dimensional grid of $P$ MPI processes. 
Inter-process communication is based on the communication-reducing 2.5D algorithm~\cite{Lazzaro:2017:IES:3093172.3093228}. 
In the tests reported in this paper, the data of the matrix multiplication $C = C + A\cdot B$ is decomposed such that it requires only the communication of the $A$ and $B$ matrix data. These communications are 
implemented with asynchronous point-to-point, MPI calls, using the MPI Funneled mode.
The local multiplication will start as soon as all the data has arrived at the 
destination process (by using a {\tt mpi\_waitall} call). The amount of communicated data by each process scales as $\mathcal{O} (1/\sqrt{P})$.

The local computation consists of pairwise multiplications of small dense matrix blocks, with dimensions 
$(m \times k)$ for $A$ blocks and $(k \times n)$ for $B$ blocks.
 It employs a cache oblivious matrix traversal to fix the order in which matrix blocks need to be computed, in order to improve memory locality. 
First, the algorithm loops over $A$ matrix row-blocks and then, for each row-block, over $B$ matrix column-blocks.
A filtering procedure is applied on the multiplication (on-the-fly filtering) of the blocks so that only blocks for which the product of their norms exceeds a given threshold will be actually multiplied. This filtering increases sparsity but also avoids performing calculations that fall below the filtering threshold, which results in a significant speed-up of the entire operation~\cite{joost1M}.
Then, the corresponding multiplications are organized in batches. 
Multiple batches can be computed in parallel on the CPU by means of OpenMP threads or alternatively executed on a GPU.  A static assignment of batches with a given $A$ matrix row-block to threads is employed in order to avoid race conditions.  
Processing the batches has to be highly efficient. For this reason specific libraries were developed that outperform vendor BLAS libraries, namely \LIBCUSMM for GPU and \LIBXSMM for CPU/KNL systems~\cite{ole, libxsmm}.

\begin{figure}
\centering
\includegraphics[scale=0.7]{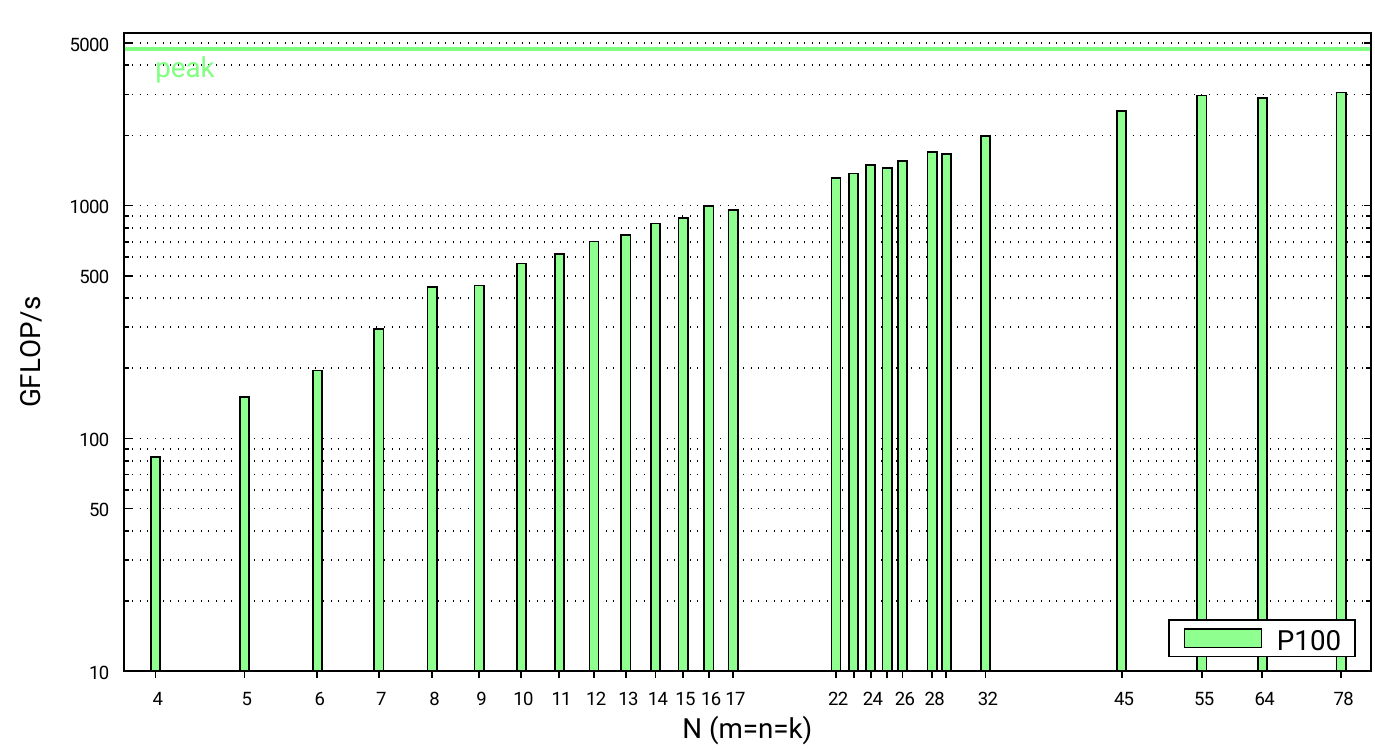}
\caption{\LIBCUSMM performance on Nvidia P100 card for selected optimized CUDA-kernels comprising $(m=n=k) = \{4,...,78\}$ block sizes. The FLOP-rates, as obtained from individual kernel launches in a mini-app that mimics \DBCSR multiplication of batches, are shown as green bars. The horizontal line refers to the double-precision peak performance of the Nvidia P100 used in our tests.}
\label{fig:cuda_kernels}
\end{figure}

For GPU execution, data is organized in such a way that the transfers between the host and the GPU are minimized. A double-buffering technique, based on CUDA streams and events, is used to maximize the occupancy of the GPU and to hide the data transfer latency.
When the GPU is fully loaded, computation may be simultaneously done on the CPU. \LIBCUSMM employs an auto-tuning framework to find optimal parameters and implementations
for each given set of block dimensions.  In this way the library is able to achieve a speedup in the range of 2--4x with respect to batched
DGEMM in cuBLAS~\cite{ole}. For Nvidias P100 we re-optimized the kernel parameter set. The performance results for the execution of blocked multiplication batches are shown in Figure~\ref{fig:cuda_kernels}. 

\LIBXSMM is a library targeting Intel Architecture for small, dense or sparse matrix multiplications, and small convolutions. The library generates executable code Just-In-Time (JIT) by assembling the instructions in-memory. All flavors of AVX extensions are supported via JIT-code, and particular emphasis is given to AVX-512. Besides redispatching generated code for every multiplication, the library can generate or dispatch the code ahead of time. This is used by \DBCSR as the block sizes of the multiplication batches are known upfront. To quantify the advantage of \LIBXSMM over vendor BLAS, we measured the performance (DP-GFLOP/s) of multiplications $C = C + A_{i} \cdot B_{i}$ with $i = 1\dots N$ such that $N$ amounts to a working set of 2\,GB (\LIBXSMM's SMM sample), and calculated the geometric mean of the performance for a series of kernels (the same as in \figurename~\ref{fig:cuda_kernels}). Streaming $A$ and $B$ matrices from memory (DDR4 or MCDRAM), and accumulating into $C$ (cached) conforms with \DBCSR's batched block multiplication. 
We measured a speedup of 2.9x for \LIBXSMM over \MKL. We have not implemented \MKL's batch-GEMM in \DBCSR and did not try MKL\_DIRECT,  but expect \LIBXSMM to maintain the advantage~\cite{libxsmm}. In absolute numbers when compared to \figurename~\ref{fig:cuda_kernels}, KNL yields higher absolute performance for smaller kernel sizes. The latter is true even when the mini-app (used to tune \LIBCUSMM) is assumed to stream $A$ and $B$ matrices from memory (rather than running hot in cache). In turn, relying on in-cache block multiplications with \LIBXSMM peaks at 1.9~TF/s (32x32 kernel).

\section{Performance Results}
\label{sec:results}

We present the results of running \DBCSR within \CPTWOK benchmark applications, representative of large-scale and long-running science runs of \CPTWOK for linear scaling calculations. Importantly, these result in
matrices with different block sizes and occupation, which affects performance and scalability.  

Timings are obtained from a \CPTWOK internal timing framework.  We did not perform any lower-level measurements of performance, such as based on hardware event counters.
We considered only the execution time of the \DBCSR multiplication part, and not any other \CPTWOK specific parts.
Results are taken as the average of 4 independent application runs, each consisting of tens of multiplications -- fluctuations are found to be less than 5\%.
Elements of the generated matrices are double precision floating point numbers.

\subsection{Single-node Performance Results}
\label{sec:singleperf}

\begin{figure}
\centering
\includegraphics[scale=0.7]{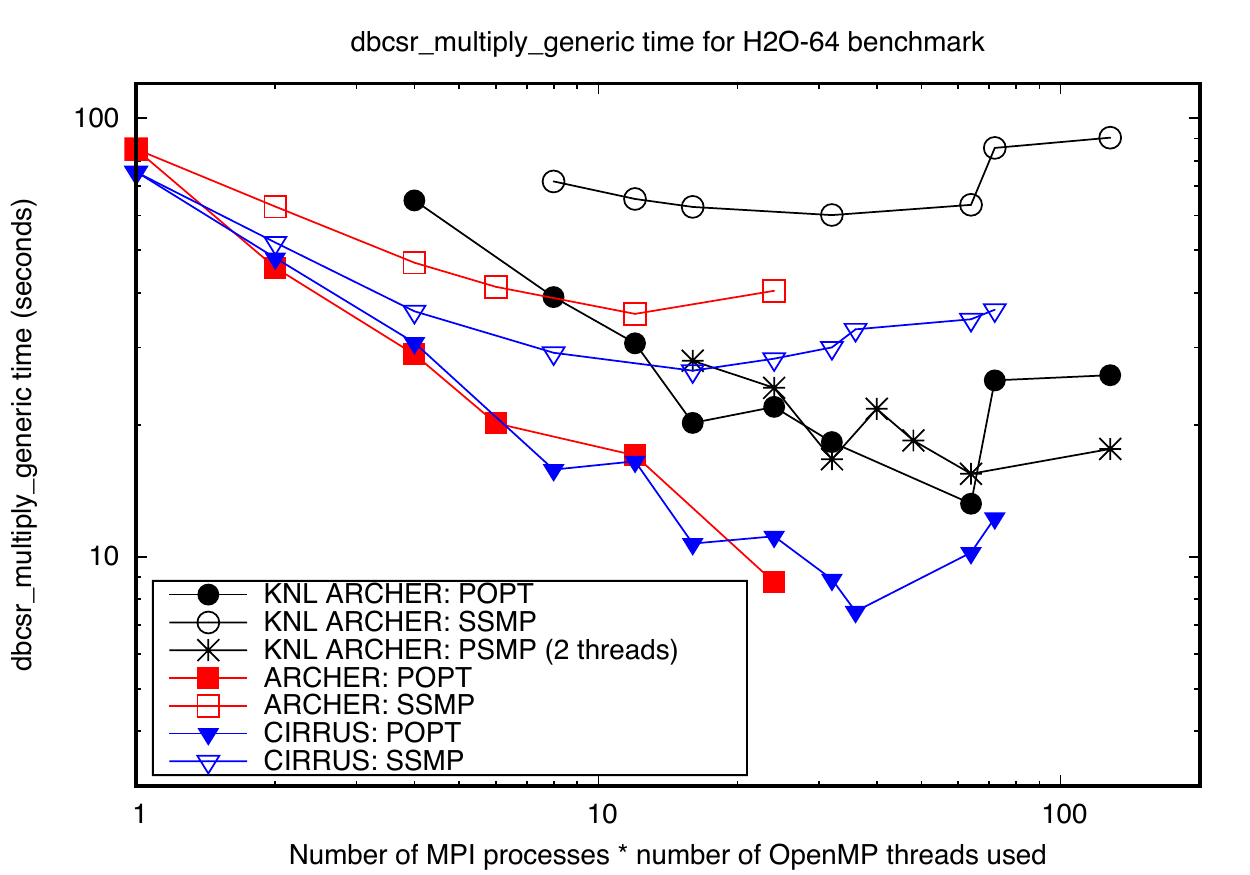}
\caption{{\tt H2O-64} performance for ARCHER, Cirrus and ARCHER-KNL on a single node execution. POPT denotes a pure MPI run, SSMP denotes pure OpenMP run and PSMP denotes a run using both MPI and OpenMP. }
\label{fig:h2o-64}
\end{figure}

We present the results of running the \CPTWOK\ {\tt H2O-64} benchmark (a small system of 64 water molecules) on the three systems:
\begin{itemize}
\item {\it ARCHER}: 4920 Cray XC30 compute nodes with Intel Xeon E5-2697 v2 (12 cores, dual-socket @ 2.7~GHz), 64~GB of RAM on 4544 nodes and 128~GB of RAM on the remaining 376 nodes. 
\item {\it Cirrus}: 280 node SGI ICE XA with Intel Xeon E5-2695 v4 (18 cores, dual-socket @ 2.1~GHz), 256~GB of RAM;
\item {\it ARCHER-KNL}: 12 Cray KNL compute nodes (64 cores Intel Xeon Phi CPU 7210 @ 1.3~GHz, 16~GB MCDRAM), 96~GB of RAM.
\end{itemize}
The goal of these tests is to get a first insight into the KNL performance. 
The results are shown in Figure~\ref{fig:h2o-64}. The block sizes for this test are combinations of $(m,n,k) = \{9, 22, 32\}$  with fully occupied matrices.
For each system we ran the benchmark in pure MPI (POPT), pure OpenMP (SSMP) and for KNL only using both MPI + OpenMP (PSMP). On the KNL we used a node with full CACHE and QUADRANT clustering and tested all possible combinations of MPI processes and OpenMP threads. We include the best PSMP result which used 2 OpenMP threads per process. 

For each system the POPT version running on a fully populated node gives the best performance with the SSMP version generally being around two times slower. On Cirrus and the ARCHER-KNL where the number of cores is greater than 64, we used hyperthreading (a maximum of 2 threads per core was tested) and it is obvious that this does not enhance the performance. The fastest result of 7.5s is obtained using 36 MPI processes on Cirrus, with the best result on ARCHER being 8.7s on 24 MPI processes and 13.2s on the ARCHER-KNL on 64 MPI processes. 

\subsection{Multiple-node Performance Results}
\label{sec:multiperf}

Tests are based on three \CPTWOK benchmarks: 
\begin{itemize}
\item \verb+S-E+: semi-empirical benchmark with 186,624 water molecules - highly sparse matrices (average occupancy 0.05\%).
\item \verb+H2O-DFT-LS+: single-point 
energy calculation with linear scaling DFT consisting of 20,736 atoms -
medium sparsity matrices (average occupancy 10\%).
\item \verb+AMORPH+: single-point 
energy calculation with linear scaling DFT consisting of 13,846 atoms - low sparsity matrices (average occupancy 70\%).
\end{itemize}
The block sizes, total number of rows/columns (all matrices are square), typical occupancy during the simulations, number of multiplications, and FLOPs executed by \DBCSR part only are reported in Table~\ref{table:DBCSR}.

\begin{table}
\centering
\caption{Block sizes, 
dimension of matrices (rows and columns), typical occupancy of the matrices, number of multiplications performed, and \DBCSR FLOPs for the three benchmarks. }
\label{table:DBCSR}
\centering
\begin{tabular}{lccc}
\hline
 & \texttt{S-E} & \texttt{H2O-DFT-LS} &  \texttt{AMORPH} \\
\hline
Block sizes $(m, n, k)$ & $6$ & $23$ & $5, 13$ \\
\# Rows/columns &  $1,119,744$ & $158,976$ & $141,212$ \\
Occupancy range (\%) & $(4 - 6) \times 10^{-2}$ & $7 - 15$ & 
                       $34 - 77$ \\
\# Multiplications & $618$ & $193$ & $187$\\
\DBCSR FLOPs ($\times 10^{12}$)   & $74$ & $4,038$ & $3,656$ \\
\hline
\end{tabular}
\end{table}

We compare the performance obtained on several systems based on Intel Xeon CPUs, Nvidia P100 GPUs, and KNL, 
hosted at the Swiss National Supercomputing Centre (CSCS):
\begin{itemize}
\item {\it Grand Tav\'e}: 164 Cray XC40 compute nodes (64 cores Intel Xeon Phi CPU 7230 @ 1.3~GHz, 16~GB MCDRAM), 96~GB of RAM; 
\item {\it Piz Daint}: this system has two partitions: 5,320 Cray XC50 hybrid compute nodes (GPU partition) with Intel Xeon E5-2690 v3 (12 cores single socket @ 2.6~GHz) 
and Nvidia Tesla P100 (16~GB High Bandwidth Memory), 64~GB of RAM; 
1,431 Cray XC40 CPU compute nodes (MC partition) with Intel Xeon E5-2695 v4 (18 cores, dual-socket @ 2.1~GHz), 64~GB of RAM.
\end{itemize}
All CPU cores have Intel Turbo and Intel Hyper Threading Technology enabled. The
latter is not used in our benchmark runs, \IE each thread runs on a single physical core. Indeed, we found that running more threads per core does not give any speed-up.
Both systems feature Cray's Aries network. 
We also found that the module {\tt craype-hugepages2M}, which enables page sizes of 2~MB, 
gives an average speed-up of 18\% for the KNL runs.

We obtained the best performance by using a single MPI rank and 12 threads per node on the GPU partition, 4 ranks and 9 threads on the MC partition, and 4 ranks and 16 threads on KNL. These configurations give the best performance of all ranks/threads in a node with a speed-up of up to 40\%. This result is an implicit consequence of the multiplication algorithm, which gives better performance for the communications of data (computation is not affected) when a minimal square number of ranks is employed~\cite{Lazzaro:2017:IES:3093172.3093228}. The total number of MPI ranks for the KNL and MC benchmarks is 4 times the ranks of the GPU ones, which implies twice as much data to communicate per rank as a consequence of the multiplication algorithm (see Section~\ref{sec:DBCSR}).

All tests on KNL are 
executed in full CACHE mode for the MCDRAM management and QUADRANT clustering mode. It is worth underlining that the entire \CPTWOK application requires a maximum of 10~GB per node, therefore it fits entirely in MCDRAM. Specific tests requiring the application to run in MCDRAM (by using FLAT mode and forcing all allocations in MCDRAM) did not show any significant speed-up in performance.

\begin{figure}
\centering
\includegraphics[scale=0.42]{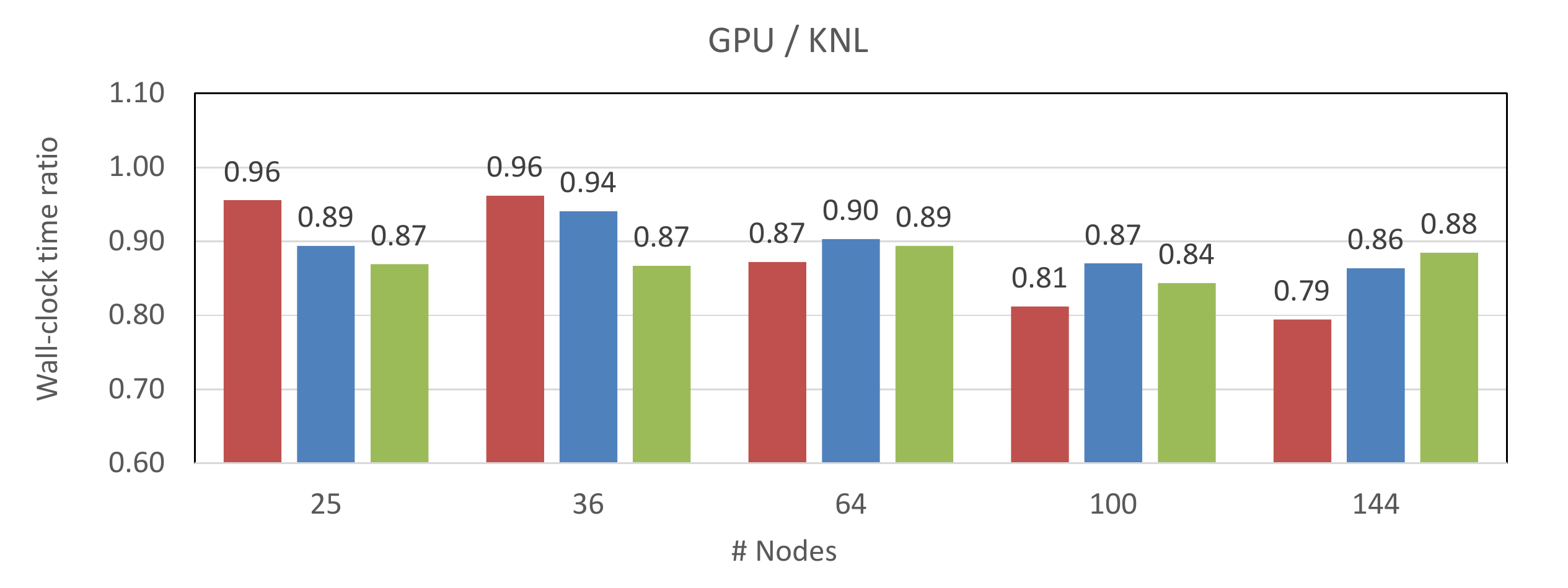}
\includegraphics[scale=0.42]{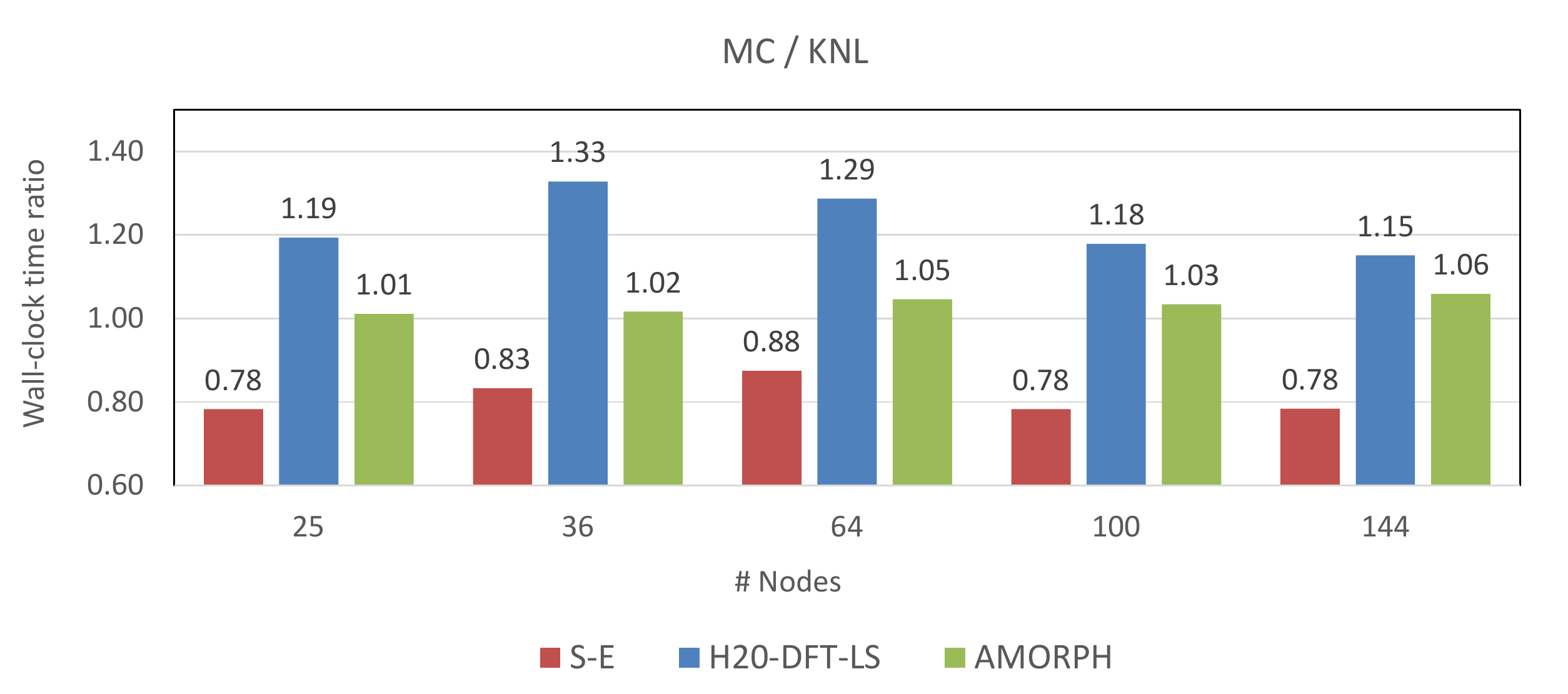}
\caption{Ratio of \DBCSR execution times between GPU (upper plot) and MC (lower plot) 
results with respect to KNL results: values greater (lower) than 1 mean that KNL executions are faster (slower).
For each number of nodes, the bars refer to (from left to right): \texttt{S-E}, \texttt{H2O-DFT-LS}, \texttt{AMORPH}.
}
\label{fig:time_comp}
\end{figure}

The \DBCSR multiplication execution times for the three systems are reported in Table~\ref{table:multi_results}. We also report the average fractions of time spent 
in the {\tt mpi\_waitall} call, used for the communication of the $A$ and $B$ matrices data,  and for the computation of the block multiplication batches. 
The time spent in the {\tt mpi\_waitall} call 
is not the full communication time for the exchange of the data, but only the part that did not overlap with computation of the 
block multiplication batches.
The remaining part of the execution time is the organization, scheduling and finalization of the matrix block multiplications. It is partially parallelized with OpenMP and is memory-bandwidth-bound.
It also includes the transfer of the $C$ matrix data between the central memory and the GPU memory~\cite{ole}.
The performance ratios are shown in Figure~\ref{fig:time_comp}. On average, we find that KNL executions are: 
\begin{itemize}
\item 11\%-14\% slower than GPU executions for the thee benchmarks;
\item 18\% slower than MC executions for \texttt{S-E};
\item 24\% and 4\% faster than MC executions for \texttt{H2O-DFT-LS} and \texttt{AMORPH}, respectively.
\end{itemize}
Although the interpretation of the data is difficult, as the algorithm is largely asynchronous, both with computation on the GPU/CPU and with communication across the network,
we can explain these results with the following observations:
\begin{enumerate}
\item From the time spent in the multiplication of batches, we see that GPU is particularly efficient for the 
\texttt{H2O-DFT-LS} and \texttt{AMORPH} benchmarks, which involve somewhat large block sizes (see Figure~\ref{fig:cuda_kernels}): on average, KNL executions are 35\% and 27\% slower than GPU executions for \texttt{H2O-DFT-LS} and \texttt{AMORPH} benchmarks, respectively, while they are 7\% faster for the \texttt{S-E} benchmark. On the other side, KNL executions are faster than MC executions for \texttt{H2O-DFT-LS} (67\%), same performance for \texttt{AMORPH}, and slower for \texttt{S-E} (13\%).
\item The time spent in the {\tt mpi\_waitall} call is directly related to the previous point, since it is the remaining time for communications that does not overlap with the computation.  
The \texttt{H2O-DFT-LS} is the most communication-bound, while the \texttt{AMORPH} is computation-bound. As expected, the communication fractions increase with the number of MPI ranks (see Section~\ref{sec:DBCSR}). Note that the GPU runs use 4 times fewer ranks than the others, therefore half as much data is communicated. 
\item As a combination of the previous two points: on average, the KNL benchmarks are between 14\%-22\% slower than GPU benchmarks and 10\%-17\% faster than MC ones. These values are partially compensated by the remaining part of execution, which scales better on the KNL and MC systems. 
\end{enumerate}
In summary, KNL gives poorer performance than GPU for the computational part when kernels are large and it requires more communication time, but it has a faster execution of the remaining part, while the opposite is true when compared to MC.  Which aspect dominates depends on the block sizes and scale of execution.

\subsection{Multi-thread Scalability}
\begin{figure}
\centering
\includegraphics[scale=0.19]{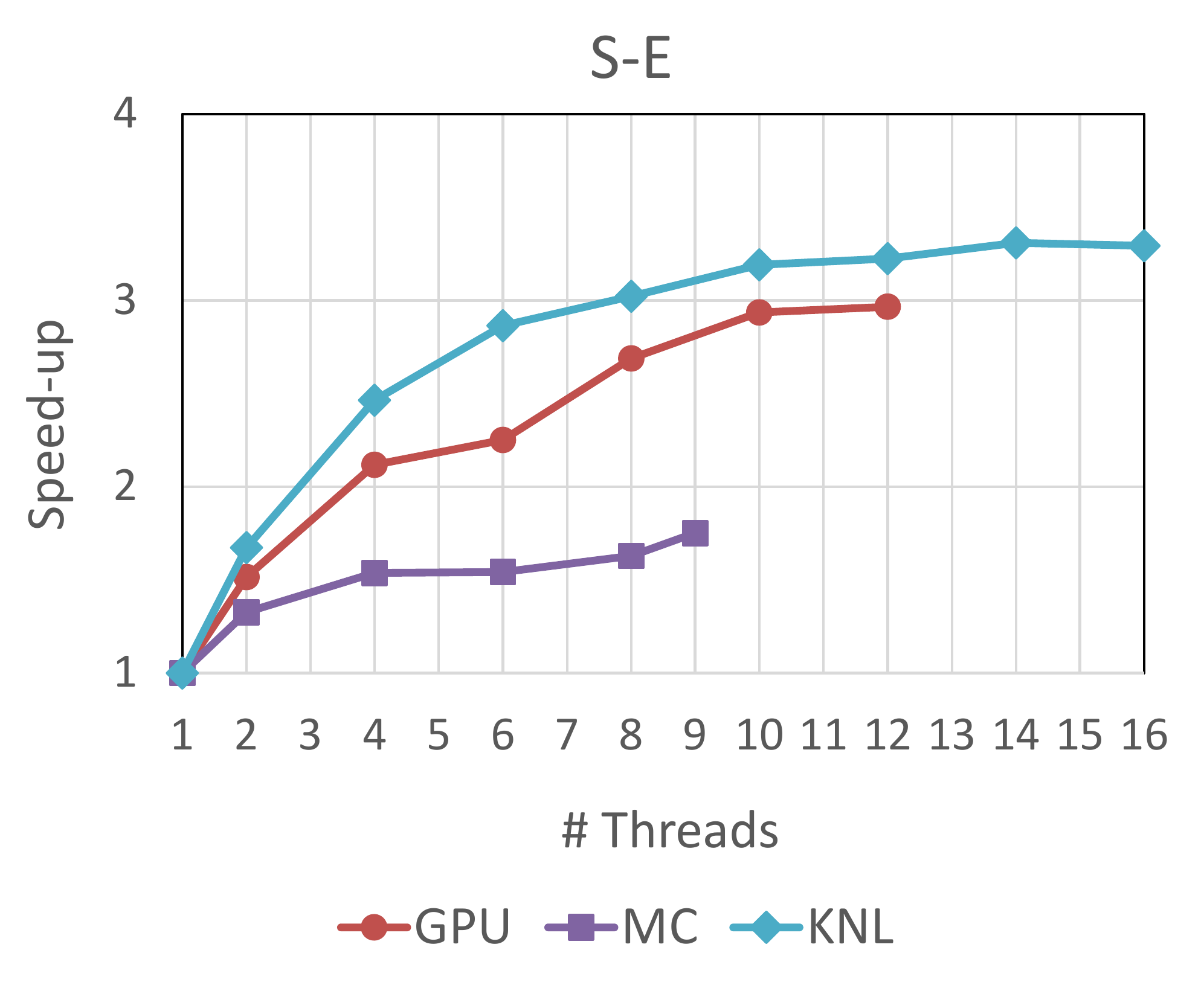}
\includegraphics[scale=0.19]{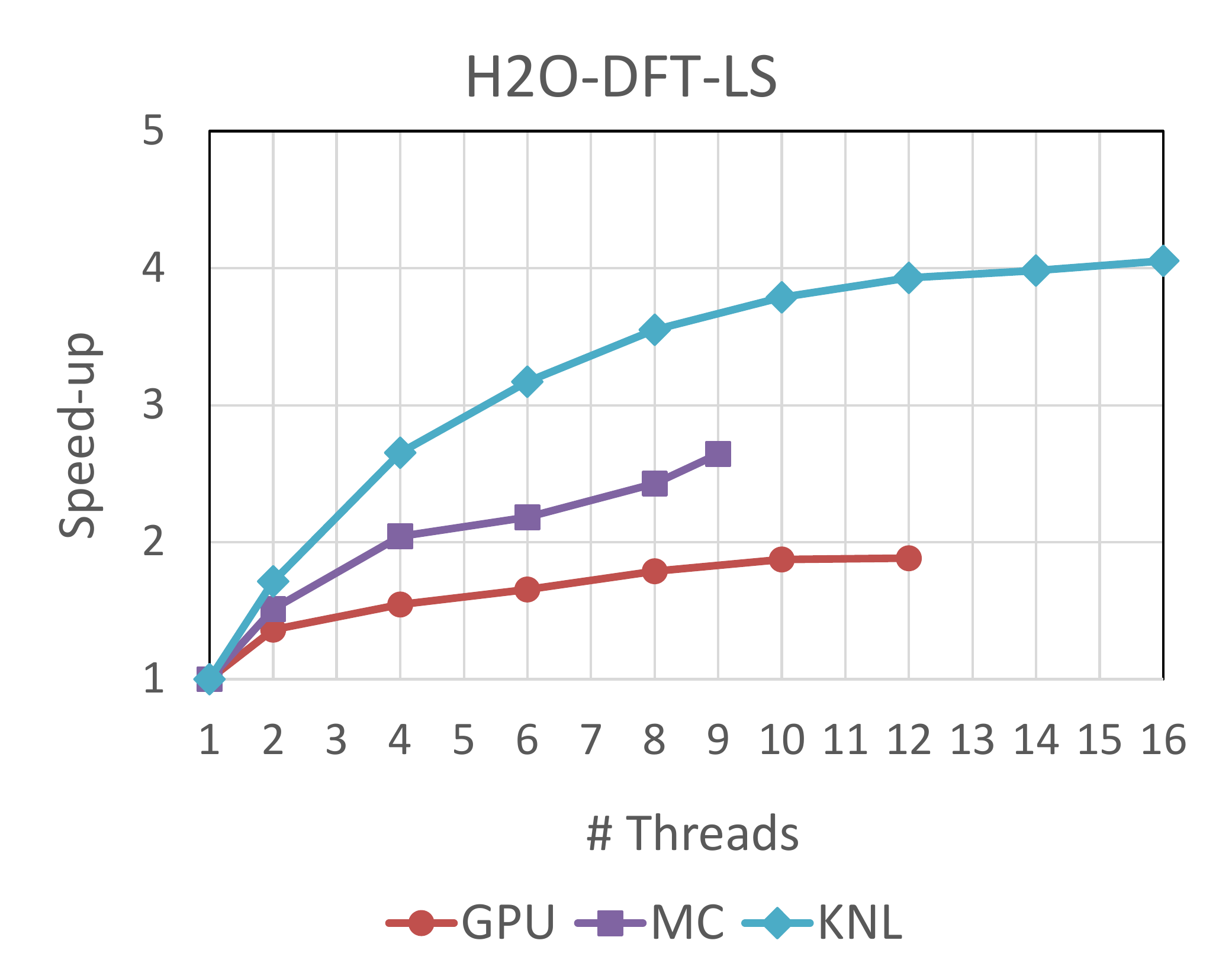}
\includegraphics[scale=0.19]{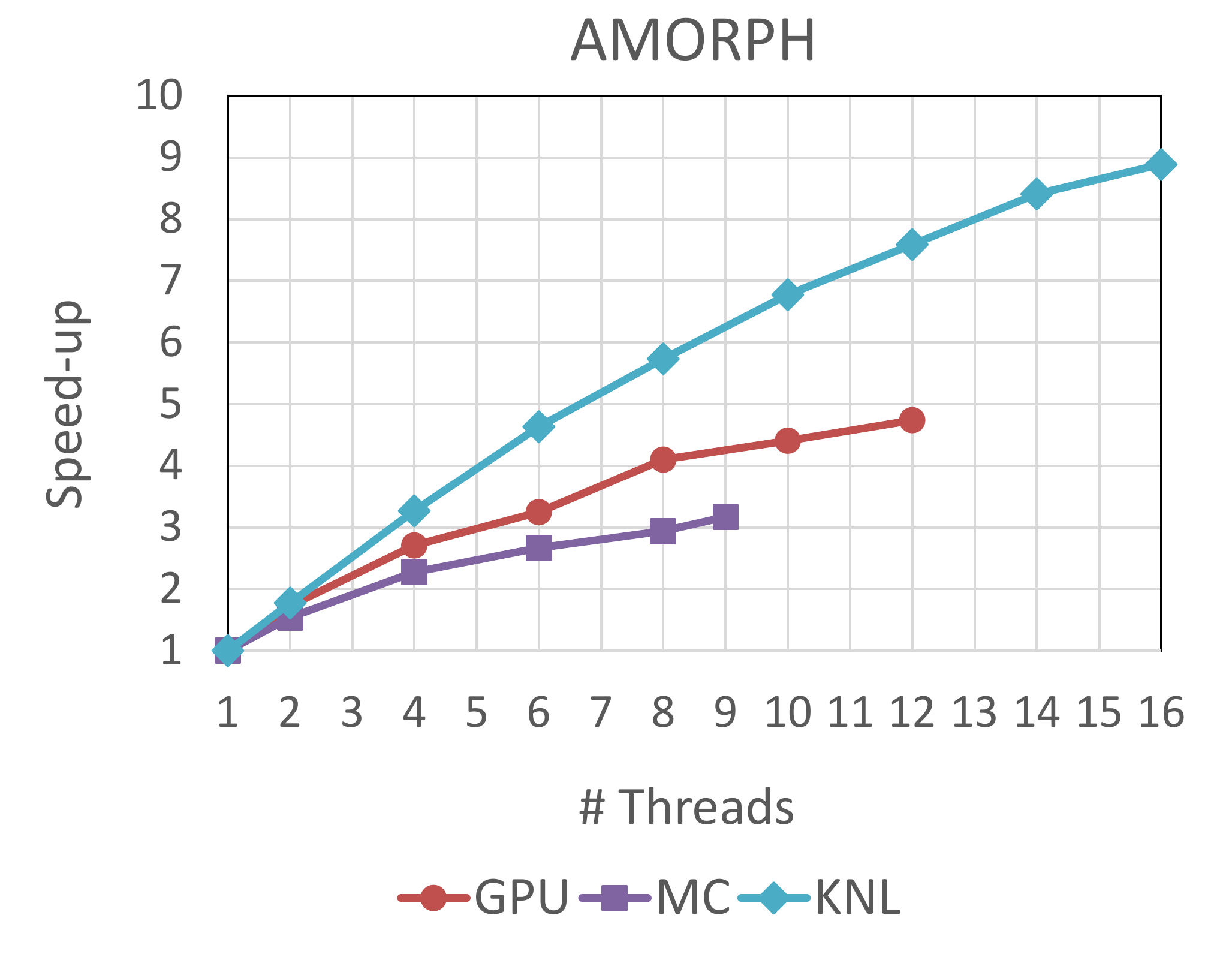}
\caption{Speed-up when varying the number of threads with respect to 
the single thread execution of the \DBCSR execution at 144 nodes for the 
\texttt{S-E}, \texttt{H2O-DFT-LS}, \texttt{AMORPH} benchmarks on the GPU, MC and KNL systems.
The number of MPI ranks is fixed for the corresponding system.
}
\label{fig:thread_speedup}
\end{figure}

We illustrate the threading scalability by considering the execution time on 144 nodes with different numbers of threads. The speedup values are shown in Figure~\ref{fig:thread_speedup}.
We can interpret these results by applying the aforementioned considerations for the execution time.  In particular, the scalability is limited by the communication time (since only the master thread handles communications) and by the initialization and finalization of the matrix multiplications, which are partially parallelized and mostly memory-bandwidth limited. Because of that, the \texttt{AMORPH} benchmark shows the best thread-scalability since it is the most computation intensive. Closer analysis of the thread timing distribution during the block multiplication batches shows some load imbalance (of up to 30\%). This is due the {\it a priori} static decomposition of the block multiplications among threads, where the load unbalance arises from the {\it a posteriori} on-the-fly filtering procedure (see Section~\ref{sec:DBCSR}). 
In the future we plan to change the algorithm to be dynamic by using OpenMP tasks. Finally, we observe that KNL scalability is always the best, due to the use of MCDRAM. Indeed, we observed a significant slowdown (between 10\%-56\%, depending on the benchmark) when performing tests where the application did not use the MCDRAM, i.e. FLAT mode forcing the allocations on DRAM.

\begin{table}
\centering
\caption{\DBCSR results (time-to-solution) for multiple nodes tests for the three benchmarks. We also report the average fractions of time spent 
in the {\tt mpi\_waitall} call  and for the computation of the block multiplication batches. 
Average timings are obtained from the values of all involed MPI ranks. 
}
\label{table:multi_results}
\centering
\begin{tabular}{c|c|ccc|ccc|ccc}
\cline{2-11} 
& \multirow{2}{*}{\# nodes}  & 
            \multicolumn{3}{c|}{\texttt{S-E}} & 
            \multicolumn{3}{c|}{\texttt{H2O-DFT-LS}} & 
            \multicolumn{3}{c}{\texttt{AMORPH}} \\
\cline{3-11} 
 &  & GPU & MC &  KNL & GPU &  MC & KNL & GPU & MC & KNL \\
\hline
\multirow{5}{2cm}{\centering Time-to-solution (seconds)} 
& 25  & 672 & 551 & 703 & 631 & 843 & 706 & 1050 & 1222 & 1208 \\ 
& 36  & 518 & 449 & 539 & 512 & 723 & 544 &  774 &  907 &  892 \\
& 64  & 361 & 362 & 414 & 374 & 533 & 414 &  493 &  577 &  551 \\
& 100 & 264 & 255 & 325 & 264 & 358 & 304 &  342 &  419 &  405 \\
& 144 & 218 & 215 & 274 & 231 & 308 & 267 &  268 &  321 &  303 \\
\hline
\multirow{5}{2cm}{\centering Average time {\tt mpi\_waitall} (\%)} 
& 25  & 18 & 38 & 25 & 28 & 22 & 31 & 2  & 3  & 2  \\
& 36  & 21 & 42 & 26 & 32 & 27 & 33 & 3  & 5  & 3  \\
& 64  & 29 & 50 & 32 & 45 & 38 & 45 & 6  & 11 & 5  \\
& 100 & 31 & 51 & 37 & 52 & 44 & 54 & 14 & 19 & 10 \\
& 144 & 35 & 56 & 40 & 55 & 52 & 57 & 21 & 27 & 15 \\
\hline
\multirow{5}{2cm}{\centering Average time for multiplication batches (\%) } 
& 25  & 25 & 26 & 21 & 33 & 57 & 42 & 67 & 80 & 77 \\
& 36  & 24 & 23 & 21 & 29 & 49 & 38 & 65 & 77 & 74 \\
& 64  & 22 & 17 & 18 & 21 & 37 & 30 & 60 & 71 & 73 \\
& 100 & 21 & 17 & 16 & 16 & 32 & 22 & 56 & 62 & 67 \\
& 144 & 19 & 15 & 14 & 13 & 26 & 18 & 51 & 55 & 62 \\
\hline
\end{tabular}
\end{table}

\section{Conclusions}
\label{sec:conclusions}

We found that the \DBCSR executions on Cray XC40 KNL-based systems are 11\%-14\% slower than on a hybrid Cray XC50 GPU based system with Nvidia P100 cards, at the same
number of nodes.  When compared to a Cray XC40 system equipped with dual-socket Intel Xeon CPUs, the KNL executions are of up to 24\% faster.
The best performance was obtained by configuring the KNL in full CACHE mode and QUADRANT clustering mode, without using hyperthreading.

\section*{Acknowledgments}
This work was supported by grants from the Swiss National Supercomputing Centre (CSCS) under projects S238, K02 and UZHP and received funding from the Swiss University Conference through the Platform for Advanced Scientific Computing (PASC).  We acknowledge support from the Engineering and Physical Sciences Research Council (EPSRC) grant EP/K038583/1 and the Intel Parallel Computing Centre programme. This work used the ARCHER UK National Supercomputing Service (http://www.archer.ac.uk) and EPCC's Cirrus HPC Service (https://www.epcc.ed.ac.uk/cirrus).

\bibliographystyle{unsrt}

\end{document}